\documentstyle[aps,pra,preprint,tighten]{revtex}
\begin{document}
\draft
\title
{Discrimination between nonorthogonal two-photon
polarization states\footnote{Contribution at the 8th Central-European
Workshop on Quantum Optics, April 27 - 30, 2001}}
\author{Ulrike Herzog}
\address{Institut f\"ur Physik,  Humboldt-Universit\"at zu 
    Berlin, Invalidenstrasse 110, D-10115 Berlin, Germany}
    
\maketitle

\begin{abstract}
 It is shown that generalized measurements, required for
 optimally discriminating between nonorthogonal 
 joint polarization states of two indistinguishable photons,
 can be realized with the
 help of polarization-dependent two-photon absorption and by 
 means of sum-frequency  generation.
 Optimization schemes are investigated
 with respect to minimizing the error probability in inferring the
 states, as well as with respect to maximizing the probability of
 success for unambiguous discrimination. Moreover,
 an implementation of error-minimizing discrimination 
 between $N$ symmetric single-photon states is studied. 
 The latter can be used to extract information from the inconclusive
 results occurring in  
 unambiguous discrimination
 between three symmetric two-photon polarization states.
 

\end{abstract}

\section{Introduction}
The problem of state discrimination consists in determining the
actual state of a quantum system that is prepared in an unknown
state belonging to a known finite set of given pure states.
Nonorthogonal quantum states cannot be perfectly discriminated,
and therefore optimization strategies have
been developed that minimize either the probability of errors,
or, when unambiguous discrimination is required, the probability of
getting an inconclusive result \cite{chefrev}.
For single-photon polarization
states, experiments have been performed that realize optimum
unambiguous discrimination between two
states \cite{clarke} and error-minimization for up to four states
of a specific kind \cite{clarke1}.
In this contribution we consider the discrimination between the
states characterizing the joint polarization of two photons 
travelling in a single spatial mode. The Hilbert space of these 
states is spanned by a
three-dimensional basis, corresponding to the three
possibilities of distributing two indistinguishable
photons among the two orthogonal polarization modes
of a transverse field.
Two-photon polarization states are a candidate 
for implementing a ternary 
quantum logic for quantum computation, or for applying the 
quantum cryptographic schemes that have been developed
for three-state systems \cite{bech}.
Recently three mutually orthogonal
two-photon polarization states have been experimentally
observed \cite{tseg}. 

\section{Discrimination with minimum error probability}
Given $N$ nonorthogonal quantum
states $|\psi_k\rangle$ $(k=1, 2, ...N)$ occurring with
equal a priori probability, the problem of error
minimization in state discrimination has been solved for
arbitrary $N$ under the condition that the states
are symmetric \cite{ban} which means
that each state results from its predecessor by applying a unitary
operator $\hat V$ in a cyclic way, i. e.
$|\psi_k\rangle= {\hat V}^{k-1}|\psi_1\rangle$ and
$|\psi_1\rangle= \hat{V} |\psi_N\rangle$.
In this case the maximum achievable probability $P_C$ to
infer the states correctly from a measurement, or the minimum
error probability $P_E$, respectively, obeys the equation \cite{ban}
\begin{equation}
P_C = 1 -P_E
=\frac{1}{N} \sum_{k=1}^N |\langle\mu_k |\psi_k\rangle |^2,
\hspace {0.3cm} {\rm where} \hspace{0.3cm}
 |\mu_k\rangle = \left( \sum_{k=1}^N|\psi_k \rangle
 \langle\psi_k|\right)^{-\frac{1}{2}}|\psi_k \rangle .
\label{1}
\end{equation}
In the optimized measurement scheme that realizes Eq. (\ref{1}), the
quantum system is guessed to be in the state $|\psi_k\rangle$
provided that the
state $|\mu_k\rangle$ is detected.

Let us consider $N$ specific  symmetric 
quantum states given by
\begin{equation}
|\psi_k\rangle=\sum_{l=0}^M c_l\,
{\rm e}^{{\rm i}\frac{2\pi}{N}l\,k}\,|u_l\rangle,
\hspace {0.3cm} {\rm with} \hspace {0.3cm} k=1,2,\dots N
\hspace {0.3cm} {\rm and} \hspace {0.3cm} N \geq M+1,
\label{2}
\end{equation}
where the  coefficients $c_l$ are
nonzero complex numbers and
the state vectors $|u_l\rangle$ represent a set of
$M+1$ orthonormal basis states to be specified
later.
We find the optimum detection states to be
\begin{equation}
|\mu_k\rangle= \frac{1}{\sqrt{N}}\sum_{l=0}^M
\frac{c_l}{|c_l|}\,
{\rm e}^{{\rm i}\frac{2\pi}{N}l\,k}\,|u_l\rangle
\hspace {0.3cm} {\rm with} \hspace {0.3cm}
\langle \mu_j|\mu_k\rangle_{j\neq k}
= \frac{1}{N}
 \frac{{\rm e}^{{\rm i}\frac{2\pi}{N}(k-j)(M+1)}-1}
       {{\rm e}^{{\rm i}\frac{2\pi}{N}(k-j)}-1},
\label{3}
\end{equation}
 from   
which we obtain the probability of correct guesses
$P_C= \frac{1}{N}\left( \sum_{l=0}^M |c_l|\right)^2.$
When $N = M+1$, the $N$ states defined by Eq. (\ref{2}) are 
linearly independent. In this case the detection states 
(\ref{3}) are normalized and mutually orthogonal,
the detection operators
$|\mu_k \rangle \langle\mu_k|$ therefore being
conventional projection operators.
For $N >M+1$ the states given by Eq. (\ref{2})
form an overcomplete set of linearly dependent states,
and the resulting nonorthogonal optimum
detection states are nonnormalized since
$\langle \mu_k|\mu_k\rangle= (M+1)/N$.
The positive Hermitian operators
$|\mu_k \rangle \langle\mu_k|$ satisfy the resolution
of the identity $\sum_{k=1}^N |\mu_k \rangle \langle\mu_k|=
\hat{1}$ and can be interpreted to be quantum
detection operators of a generalized measurement based on
positive-operator valued measures \cite{ban}.

In the following we restrict ourselves to the case 
$M=2$ and 
study error-minimizing 
discrimination between 
the $N$  specific symmetric two-photon
polarization states
\begin{equation}
|\psi_k\rangle=c_0 |u_0\rangle +
c_1 {\rm e}^{{\rm i}\frac{2\pi}{N} k}|u_1\rangle +
c_2 {\rm e}^{{\rm i}\frac{4\pi}{N} k}|u_2\rangle
\hspace{0.5cm}
{\rm with} \hspace{0.5cm}
 |c_2| \leq |c_0|,|c_1|,
\label{5}
\end{equation}
where $k=1,\ldots N$. The basis states are now defined by
\begin{equation}
|u_0\rangle  \equiv  |2,0\rangle =
    \frac{\hat{a}_1^{\dag^2}}{\sqrt{2}}|0\rangle,\hspace{0.2cm}
|u_1\rangle  =  |1,1\rangle\rangle =
              \hat{a}_1^{\dag}\hat{a}_2^{\dag}|0\rangle,\hspace{0.2cm}
{\rm and}\hspace{0.2cm}
 |u_2\rangle  \equiv  |0,2\rangle =
    \frac{\hat{a}_2^{\dag^2}}{\sqrt{2}}|0\rangle.
\label{6}
\end{equation}
Here the photon creation operators
$\hat{a}_1^{\dagger}$ and $\hat{a}_2^{\dag}$ refer to any two
mutually orthogonal polarization modes of the field.
An interesting special case arises
when the $k$th state is created by the operator
$\hat{b}_k^{\dagger}=(\hat{a}_1^{\dagger} +
{\rm e}^{{\rm i}\frac{2\pi}{N}k}\, \hat{a}_2^{\dagger})/\sqrt{2}$.
The two-photon states
$ 2^{-1/2}(\hat{b}_k^{\dagger})^2|0\rangle$
are then given by Eq. (\ref{5}) with
$c_0=c_2=1/2$ and $c_1=1/\sqrt{2}$, yielding
the value $P_C = (3+2\sqrt{2})/(2N)$ for the 
maximum achievable  probability of correct guesses. 
This is value 
is larger than  the result $P_C = 2/N$
following for  the corresponding single-photon 
states $ \hat{b}_k^{\dagger}|0\rangle = (|1,0\rangle +
{\rm e}^{{\rm i}\frac{2\pi}{N}k}\,|0,1\rangle )/\sqrt{2}$.

For a physical implementation of state discrimination with
minimum error probability, one could make use of two-photon
absorption.
Let us consider an atom that is pumped into a coherent
superposition of three degenerate lower energy
states $|g_0\rangle,
|g_1\rangle$, and $|g_2\rangle$, having the magnetic quantum numbers
$m=-2, m=0$, and $m=+2$, respectively. By two-photon
transitions, the lower levels are
assumed to be connected to an excited state $|e\rangle$ with $m=0$.
For simplicity, we write the interaction Hamiltonian as
$H=\hbar\eta \left(\hat{a}_1^2|e\rangle\langle g_0|  +
            \sqrt{2}\hat{a}_1 \hat{a}_2|e\rangle\langle g_1|  +
 \hat{a}_2^2|e\rangle\langle g_2| \right) + H. A., $
where $\eta$ is real and denotes the atom-field coupling constant.
When the orientation of the atomic quantization axis coincides with
the direction of wave propagation, $\hat{a}_1$ and
$\hat{a}_2$ refer to right-handed and left-handed circular
polarization, respectively.
We start from an initial atomic superposition state
$|\chi\rangle=\sum_{l=0}^2 \alpha_l |g_l\rangle$
and from an initial two-photon polarization state
of the field $|\psi\rangle=\sum_{l=0}^2 \beta_l |u_l\rangle$,
where the basis states are given by Eq. (\ref{6})
and refer to circular polarization.
By calculating the combined atom-field state
$|\Phi_{tot}(t)\rangle={\rm exp}(-\frac{i}{\hbar}H t)
                      ( |\chi\rangle |\psi\rangle )$,
we obtain 
after averaging with respect to the interaction time $t$
the time-independent single-atom excitation probability
\cite{herzog}
\begin{equation}
\overline{p}_e = \Gamma \int_0^{\infty} dt\,
 {\rm e}^{-\Gamma t}
 {\rm Tr}|\langle e|\Phi_{tot}(t)\rangle|^2=
 \frac{2\eta^2}{\Gamma^2+12\eta^2}
 \left|\sum_{l=0}^2 \alpha_l \beta_l\right|^2
 = \frac{2\eta^2}{\Gamma^2+12\eta^2}\
  |\langle \mu| \psi\rangle|^2.
\label{8}
\end{equation}
 Here the trace has been performed over the 
field and we  introduced the notation
$|\mu \rangle = \sum_{l=0}^2 \alpha_l^{\ast} |u_l\rangle$.
Provided that $N$ kinds of atoms,
labelled by the index $k$, are prepared
in superposition states
$|\chi_k\rangle=\sum_{l=0}^2 \alpha_l^{(k)} |g_l\rangle$
designed in such a way that
$\alpha_l^{(k)}
= \frac{1}{\sqrt{N}}\frac{c_l}{|c_l|}\,
{\rm e}^{-{\rm i}\frac{2\pi}{N}kl}$,
the respective excitation probabilities are
proportional to $|\langle\mu_k|\psi\rangle|^2$,
where $|\mu_k\rangle$ is defined by
Eq. (\ref{3}) with $N=2$.
Therefore the atomic superposition states 
represent the detection states
of the error-minimizing measurement scheme
if it is possible to observe,
e. g. by fluorescence detection,
which kind of atom has been excited.
For $N=3$, when the optimum detection states 
are mutually orthogonal,
one could use three
separate gas cells being traversed one after the other 
by the unknown two-photon polarization 
states and being each filled
with a sufficiently large number of atoms 
prepared in the realization of a different 
detection state.

For later use we still consider  the specific
symmetric single-photon states defined by 
$|\xi_k\rangle= c_0 |1,0\,\rangle
+c_1{\rm e}^{{\rm i}\frac{2\pi}{N}k}\,|0,1\rangle$
with $k=1,\ldots N$.
In this case 
error-minimizing state discrimination 
can be achieved by inferring the unknown state
to be the state $|\xi_k\rangle$ provided that the
photon is detected at the $k$th output port
of a lossles linear optical network having
$N$ input ports and $N$ output ports.
For this purpose, the two modes
have to be directed into separate input ports of the
optical multiport.
According to Eq. (\ref{1}), the latter has to be constructed in
such a way that, for any input state $|\xi_k\rangle$,
the probability $|d_j^{(out)}|^2$
that the photon exits at the output port $j$ is given by
\begin{equation}
|d_j^{(out)}|^2= |\langle\mu_j|\xi_k\rangle|^2
 =\frac{1}{N}\left[1+2|c_0||c_1|
  \cos\frac{2\pi(k-j)}{N}\right].
\label{10}
\end{equation}
Here $|\mu_j\rangle$ has been determined
from Eq. (\ref{3}) with $M=2$, using the basis states
$|1,0\rangle$ and $|0,1\rangle$.
In order to find the unitary transformation matrix $U$
characterizing a specific multiport that implements
error minimization,
we make use of the relation
$d_j^{(out)} = \sum_{r=1}^N U_{jr} d_r^{(in)}$
that connects the single-photon input and output 
probability amplitudes, or 
the classical fields, respectively.
With  $d_1^{(in)} = c_0$,
$d_2^{(in)} = c_1{\rm e}^{{\rm i}\frac{2\pi}{N}k}$, and
$d_r^{(in)} = 0$ \, ($r\geq 3$)\
the required relation (\ref{10}) is fulfilled 
provided that, for $j=1,\ldots N$,
\begin{equation}
U_{j1} = \frac{1}{\sqrt{N}}
        {\rm e}^{i({\rm Arg}\,c_1-{\rm Arg}\,c_0)}
\hspace{0.2cm}{\rm and}\hspace{0.2cm}
U_{jr}  =  \frac{1}{\sqrt{N}} {\rm e}^{-i\frac{2\pi}{N}j(r-1)}
\hspace{0.2cm} 
{\rm for}\hspace{0.2cm}r=2, \dots N.
\label{11}
\end{equation}
Once the transformation matrix is known, the desired
linear optical multiport can be constructed using beam 
splitters and phase shifters. This has been  
used recently for proposing an implementation of 
optimum unambiguous
discrimination between single-photon 
 states in which the photon is divided among 
more than two input modes \cite{sun}.

\section{Optimum unambiguous discrimination}
It has been proved that
unambiguous state discrimination, with a certain probability 
of success, is possible if and only if
the states are linearly independent \cite{chef1}, and that
$N$ linearly independent and symmetric states
can always be written in the form (\ref{2})
with $M=N-1$ and properly chosen basis states \cite{chef2}.
Provided that these states occur with equal
a priori probability, the maximum probability
of success has been derived to
be $P_D = N\,{\rm min}\, |c_l|^2$ \cite{chef2}.
Here ${\rm min}\,|c_l|^2$ is the smallest square modulus
arising from any of the coefficients $c_l$
that occur in Eq. (\ref{2}).
We mention that $P_D < P_C $ unless the states are orthogonal.
The optimum value $P_D$ can be achieved in a generalized 
measurement consisting of a two-step procedure \cite{chef2}:
First the given set of
nonorthogonal states has to be
transformed into a set of orthogonal ones by means of a
suitable outcome-conditioned nonunitary transformation.
By this operation, a quantum system prepared in one of
the given nonorthogonal states $|\psi_k\rangle$
will, with a certain probability of success,
be transformed into the corresponding member of a set of
orthogonal states $|\tilde{\psi}_k\rangle$.
In a second step the resulting orthogonal states can be
perfectly discriminated by a conventional
quantum mechanical projection measurement.
Here we are interested in the physical mechanisms
that enable optimum unambiguous discrimination between the
three linearly independent two-photon polarization states
given by Eq. (\ref{5}) with $N = 3$, yielding the
optimum value $ P_D = 3\,|c_2|^2 $.

First we discuss state orthogonalization via 
polarization-dependent two-photon absorption.
We assume that atoms are prepared in such a way that they
can perform two-photon absorbing transitions provided that the
photons belong to prescribed polarization modes $i$ and $j$ with
$i,j = 1,2$, and that single-photon absorption is negligible.
With $\rho$ being the reduced density operator of the radiation field,
the master equation describing the absorption process reads
$\dot{\rho}=-\frac{\gamma_{ij}}{2}(
\hat{a}_i^{\dag}\hat{a}_j^{\dag}\hat{a}_{i}\hat{a}_j\rho
 +\rho \hat{a}_i^{\dag}\hat{a}_j^{\dag}\hat{a}_i\hat{a}_j )+ S\rho.$
Here we used the abbreviation
$S\rho=\gamma_{ij} \hat{a}_i\hat{a}_j
    \rho \hat{a}_i^{\dag}\hat{a}_j^{\dag}$, and
$\gamma_{11}, \gamma_{22}$, and $\gamma_{12}=\gamma_{21}$
are the respective two-photon absorption constants.
The superoperator $S$ can be interpreted to be
a jump operator responsible for jump-like changes of the
density operator $\rho$ due to the absorption of two photons.
Under the condition that no photon is absorbed,
the evolution of the radiation field is described by
a nonnormalized conditioned density operator
$\tilde{\rho}$ obeying the evolution equation
that ensues from the master equation  when the jump-operator
term is omitted \cite{car}.
Therefore, if the radiation field is initially in the pure state
$|\psi\rangle$, the conditioned state remains pure and
is given by
$|\tilde{\psi}(t)\rangle = {\rm exp}(-\frac{\gamma_{ij}}{2}
\hat{a}_i^{\dag}\hat{a}_j^{\dag}\hat{a}_i\hat{a}_j t)
|\psi\rangle$. The probability that no two-photon absorption
process occurs is equal to
$\langle\tilde{\psi}(t)|\tilde{\psi}(t)\rangle$ \cite{car}.
We want to perform unambiguous state discrimination for
the three two-photon polarization states
$|\psi_k\rangle$ ($k=1,2,3$) that are defined by Eq. (\ref{5})
with $N=3$.
Let us assume that the states interact
during a time interval $T_0$ with a two-photon absorbing
medium for which only $\gamma_{11}$ is different from zero,
and that after the end of this interaction
(or before its beginning) the states interact
during a time interval $T_1$ with
another two-photon absorber for which only $\gamma_{12}$
differs from zero.
When both types of interaction are finished, and on
the condition that no absorption has occurred,
the incoming state $|\psi_k\rangle$ is transformed into
the nonnormalized state
$|\tilde{\psi}_k^{\prime}\rangle=
c_0 {\rm e}^{-\gamma_{11}T_0}|u_0\rangle +
c_1 {\rm e}^{-\frac{\gamma_{12}}{2}T_1}
      {\rm e}^{{\rm i}\frac{2\pi}{3}k}|u_1\rangle +
c_2 {\rm e}^{{\rm i}\frac{4\pi}{3}k}|u_2\rangle.$
To achieve orthogonalization, the interaction times have to be
adjusted in such a way that
${\rm exp}(-\gamma_{11}T_0) = |c_2|/|c_0|$ and
${\rm exp}(-\gamma_{12}T_1/2) = |c_2|/|c_1|$. This yields
the set of nonnormalized no-absorption-conditioned state vectors
\begin{equation}
|\tilde{\psi}_k\rangle=|c_2|\left(\frac{c_0}{|c_0|}|u_0\rangle +
 \frac{c_1}{|c_1|}{\rm e}^{{\rm i}\frac{2\pi}{3}k}|u_1\rangle +
 \frac{c_2}{|c_2|}{\rm e}^{{\rm i}\frac{4\pi}{3}k}|u_2\rangle\right)
\label{13}
\end{equation}
with $k=1,2,3$, which are
mutually orthogonal and can be perfectly discriminated.
The probability for successful discrimination,
i. e. for the absence of absorption, is
found to be $\langle\tilde{\psi}_k|\tilde{\psi}_k\rangle=3|c_2|^2$
which is equal to the optimum achievable value $P_D$.
When the two incoming photons are absorbed,
an inconclusive result is obtained.

If, instead of two-photon absorption, sum-frequency generation is 
used for state orthogonalization,  
the inconclusive results can be detected as well.
The interaction Hamiltonian reads
$H_{ij}=i\hbar \frac{\kappa_{ij}}{2}
              (\hat{a}_i^{\dag}\hat{a}_j^{\dag}\hat{b}_{ij} -
                      \hat{a}_i\hat{a}_j\hat{b}_{ij}^{\dag})$
with $i,j=1,2$, where $\hat{b}_{ij}^{\dag}$ 
is the creation operator for a
photon in the corresponding up-converted mode.
When   
$\hat{a}_1^{\dag}$ and $\hat{a}_2^{\dag}$  
refer to horizontally and vertically linearly polarized light,
respectively, 
the coupling constants
$\kappa_{11}$ and $\kappa_{22}$, on the one hand, and
$\kappa_{12}=\kappa_{21}$ on the other, 
correspond to type-I and type-II
nonlinear crystals. For the purpose of state orthogonalization
 we assume that the two-photon polarization 
states $|\psi_k\rangle$, 
defined by Eq. (\ref{5}) with $N=3$,
interact during a time interval $T_0$ with a type-I
crystal with $\kappa_{11} \neq 0$ and during a time interval $T_1$
with a type-II crystal.
The interaction times are now required to obey the equations
$\cos(\kappa_{11}T_0/\sqrt{2}) = |c_2|/|c_0|$ and
$\cos(\kappa_{12}T_1/2) = |c_2|/|c_1|.$
When the interaction is completed, 
the state vector of the enlarged system, including the
up-converted modes labelled by $A$ and $B$ for the type-I and type-II
crystal, respectively, takes the form 
\begin{equation}
|\Psi_k^{tot}\rangle
 =  {\rm e}^{-\frac{i}{\hbar}H_{11}T_0}
{\rm e}^{-\frac{i}{\hbar}H_{12}T_1}
|0\rangle_A\,|0\rangle_B\,|\psi_k\rangle
  =   |0\rangle_A\,|0\rangle_B\,|\tilde{\psi}_k\rangle +
  |\tilde{\xi}_k\rangle\,|0\rangle,
\label{16}
\end{equation}
where $|0\rangle$ is the vacuum state of the fundamental mode,
$|\tilde{\psi}_k\rangle$ is given by Eq. (\ref{13}), and
\begin{equation}
|\tilde{\xi}_k\rangle =
 \sqrt{|c_0|^2-|c_2|^2}\,|1\rangle_A\,|0\rangle_B
 +\sqrt{|c_1|^2-|c_2|^2}\,
     {\rm e}^{i\frac{2\pi}{3}k}\,|0\rangle_A\,|1\rangle_B.
\label{17}
\end{equation}
Provided that no sum-frequency generation occurs, the state
$|\psi_k\rangle$ is transformed into the conditioned state
$\langle 0|_A \langle 0|_B
|\Psi_k^{tot}\rangle = |\tilde{\psi}_k\rangle$, 
belonging to the set of mutually orthogonal 
states that enable optimum unambiguous discrimination.
On the other hand,
when sum-frequency generation takes place, 
the up-converted photon is found
to be in the conditioned single-photon superposition state
$|\tilde{\xi}_k\rangle$. This happens with the 
$k$-independent probability of inconclusive results  
$\langle\tilde{\xi}_k|\tilde{\xi}_k\rangle = 1-3|c_2|^2$.
The three states $|\tilde{\xi}_k\rangle$  
refer to a two-dimensional basis and are therefore
linearly dependent, rendering unambiguous 
discrimination impossible.
However, if both $|c_0|$ and $|c_1|$  are
larger than $|c_2|$, an inconclusive result still contains
information about the original state which 
can be
extracted in an optimum way 
by applying an error-minimizing discrimination 
scheme \cite{chefrev}. 
Upon normalization, the single-photon states (\ref{17})
exactly correspond to the states 
that can be discriminated with minimum error probability
using a linear multiport characterized by the transformation
matrix given by  Eq. (\ref{11})
with $N=3$. 
Hence, in order to infer the up-converted single-photon state
and thus also the original state,
 the up-converted modes
$A$ and $B$ have to be directed into two input ports
of the specific multiport.

Interestingly, except for normalization, 
the set of orthogonal states 
$\{|\tilde{\psi}_k\rangle\}$, into
which the nonorthogonal states have to be 
transformed for optimum unambiguous discrimination,
is equivalent to the set of detection
states $\{|\mu_k\rangle\}$ enabling error-minimizing 
discrimination, as can be seen by comparing 
Eq. (\ref{13}) and Eq. (\ref{3}) with $N=3$ and $M=2$. 
By generalization, we find this equivalence to be valid for linearly
indepent symmetric states in an arbitrary dimensional Hilbert space.
Finally we note that it is planned to extend the 
investigations
in order to study the discrimination between nonorthogonal  
joint polarization states of $M$ indistinguishable photons.                                            
In this case 
optimum unambiguous discrimination between up 
to $M+1$ states 
could be implemented if the necessary $M$-photon interaction processes 
were sufficiently efficient.

\end{document}